\documentclass[preprint,nofootinbib]{revtex4}
\usepackage{amssymb}
\usepackage{amsfonts}
\usepackage{amsmath}
\usepackage{graphicx}

\newcommand{\beq}{\begin{equation}}
\newcommand{\eeq}{\end{equation}}
\newcommand{\bea}{\begin{eqnarray}}
\newcommand{\eea}{\end{eqnarray}}
\begin{document}

\title{Scalar solitons and the microscopic entropy \\
of hairy black holes in three dimensions}
\author{Francisco Correa$^{1}$, Cristi\'{a}n Mart\'{\i}nez$^{1,2}$, Ricardo
Troncoso$^{1,2}$}
\affiliation{$^{1}$Centro de Estudios Cient\'{\i}ficos (CECS), Casilla 1469, Valdivia,
Chile }
\affiliation{$^{2}$Centro de Ingenier\'{\i}a de la Innovaci\'{o}n del CECS (CIN),
Valdivia, Chile}
\preprint{CECS-PHY-10/10}
\email{correa, martinez, troncoso@cecs.cl}

\begin{abstract}
General Relativity coupled to a self-interacting scalar field in three
dimensions is shown to admit exact analytic soliton solutions, such that the
metric and the scalar field are regular everywhere. Since the scalar field
acquires slow fall-off at infinity, the soliton describes an asymptotically
AdS spacetime in a relaxed sense as compared with the one of Brown and
Henneaux. Nevertheless, the asymptotic symmetry group remains to be the
conformal group, and the algebra of the canonical generators possesses the
standard central extension. For this class of asymptotic behavior, the
theory also admits hairy black holes which raises some puzzles concerning an
holographic derivation of their entropy\textit{\ \`{a} la }Strominger. Since
the soliton is devoid of integration constants, it has a fixed (negative)
mass, and it can be naturally regarded as the ground state of the
\textquotedblleft hairy sector\textquotedblright, for which the scalar
field is switched on. This assumption allows to exactly reproduce the
semiclassical hairy black hole entropy from the asymptotic growth of the
number of states by means of Cardy formula. Particularly useful is
expressing the asymptotic growth of the number of states only in terms of
the spectrum of the Virasoro operators without making any explicit reference
to the central charges.
\end{abstract}

\maketitle

\section{Introduction}

Scalar fields propagating on a fixed AdS background reveal several
interesting features. As shown during the 80's, their energy turns out to be
positive provided the squared mass is bounded from below by a negative
quantity, according to%
\begin{equation*}
m^{2}\geq m_{\ast }^{2}:=-\frac{\left( d-1\right) ^{2}}{4l^{2}}\ .
\end{equation*}%
This is known as the Breitenlohner-Freedman bound \cite%
{Breitenlohner-Freedmann, Mezin-Townsend}, which guarantees the stability of
an AdS spacetime of radius $l$ against scalar field perturbations.
Particularly interesting is the case of scalar fields whose mass is within
the range%
\begin{equation}
m_{\ast }^{2}\leq m^{2}<m_{\ast }^{2}+\frac{1}{l^{2}}\ ,  \label{bounds}
\end{equation}%
since they are able to acquire slow fall-off at infinity. In turn, this
generates a strong back reaction of the metric in the asymptotic region,
such that in certain cases they cannot be treated as a probe \cite{HMTZ-2+1,
HMTZ-Log, HMTZ-D} (see also \cite{Hertog-Maeda, Amsel-Marolf}). As a
consequence, the asymptotic behavior of the metric has to be relaxed as
compared with the standard one \cite{Brown-Henneaux, Henneaux-Teitelboim,
Henneaux-D} for a localized distribution of matter. This has the effect of
enlarging the space of physically admissible configurations, bringing in new
classes of solutions including hairy black holes and solitons \cite%
{HMTZ-2+1, MTZ1, MTZ2, MTZ3, MTZ4, MTZ5, MTZ6}. Finding exact analytic
solutions that circumvent no hair theorems is not an easy task\footnote{%
Numerical solutions have also been found in \cite{N1, N2, N3, N4, N5, N6}. The case
of minimally coupled scalar fields with electric charge has attracted much
recent attention concerning holographic superconductivity \cite{HC} (for
good reviews see, e.g. \cite{reviews}). In this context, hairy black hole
solutions have also been found numerically \cite{Manga}.}. One of the
simplest and instructive examples where this can be achieved corresponds to
General Relativity with a minimally coupled self-interacting scalar field in
three dimensions \cite{HMTZ-2+1}. The action is given by%
\begin{equation}
I[g_{\mu \nu },\phi ]=\frac{1}{\pi G}\int d^{3}x\sqrt{-g}\left[ \frac{R}{16}-%
\frac{1}{2}(\nabla \phi )^{2}-V(\phi )\right] \;,  \label{Action}
\end{equation}%
with the following self-interaction potential%
\begin{equation}
V(\phi )=-\frac{1}{8l^{2}}\left( \cosh ^{6}\phi +\nu \sinh ^{6}\phi \right)
\;,  \label{Potential}
\end{equation}%
having a global maximum at $\phi =0$, such that $V(0)=-\frac{1}{8l^{2}}$,
and a mass term given by $m^{2}=V^{\prime \prime }(0)=-\frac{3}{4l^{2}}$,
which lies within the range (\ref{bounds}). This potential has a simple
interpretation in the conformal (Jordan) frame (see Appendix \ref{a1}).

\section{Hairy black holes and relaxed AdS asymptotics}

\label{HBHs}

When the self-interaction parameter fullfills $\nu \geq -1$, apart from the BTZ black hole \cite{BTZ, BHTZ}, which is a solution of this
action in vacuum, i.e., with a vanishing scalar field, the field equations
admit an exact, static and circularly symmetric hairy black hole \cite%
{HMTZ-2+1}. The solution possesses a non trivial scalar field given by 
\begin{equation}
\phi (r)=\mathrm{arctanh}\sqrt{\frac{B}{H(r)+B}}\;,  \label{scalar bh}
\end{equation}%
where\ $H(r)=\frac{1}{2}\left( r+\sqrt{r^{2}+4Br}\right) $, and the metric
reads 
\begin{equation}
ds^{2}=-\left( \frac{H}{H+B}\right) ^{2}F(r)dt^{2}+\left( \frac{H+B}{H+2B}%
\right) ^{2}\frac{dr^{2}}{F(r)}+r^{2}d\varphi ^{2}\;,  \label{metric bh}
\end{equation}%
with $F(r)=\frac{H^{2}}{l^{2}}-(1+\nu )\left( \frac{3B^{2}}{l^{2}}+\frac{%
2B^{3}}{l^{2}H}\right) $, and the coordinates range as $-\infty <t<\infty $, 
$r>0$, $0\leq \varphi <2\pi $. Note that the hairy black hole is
well-defined for $\nu \geqslant -1$, and it depends on a single non-negative
integration constant $B$, such that the scalar field is regular everywhere.
The curvature singularity at the origin is enclosed by an event horizon
located at%
\begin{equation}
r_{+}=B\Theta _{\nu }\,,
\end{equation}%
where 
\begin{equation}
\Theta _{\nu }=2(z\bar{z})^{\frac{2}{3}}\frac{z^{\frac{2}{3}}-\bar{z}^{\frac{%
2}{3}}}{z-\bar{z}},~\text{\textrm{with}}\quad z=1+i\sqrt{\nu },
\label{Schuster}
\end{equation}%
is a function of the parameter $\nu $ appearing in the potential. The
Hawking temperature and the mass, are given by%
\begin{equation}
T=\frac{3B(1+\nu )}{2\pi l^{2}\Theta _{\nu }},~\text{\textrm{and}}\quad M=%
\frac{3B^{2}(1+\nu )}{8Gl^{2}},  \label{Temperature and Mass}
\end{equation}%
respectively\footnote{%
The mass of the hairy black hole has also been obtained following different
approaches in Refs. \cite{glenn, glenn2, GMT, Clement, BanadosThiesen}.},
and the entropy reads%
\begin{equation}
S=\frac{A}{4G}=\frac{\pi r_{+}}{2G}\ .  \label{Entropy}
\end{equation}

As also shown in \cite{HMTZ-2+1}, the hairy black hole describes an
asymptotically AdS spacetime in a relaxed sense as compared with the
standard one of Brown and Henneaux \cite{Brown-Henneaux}. Indeed, the
asymptotic behavior of the hairy black hole belongs to the following class:%
\begin{equation}
\phi =\frac{\chi }{r^{1/2}}+\alpha \frac{\chi ^{3}}{r^{3/2}}+O(r^{-5/2})
\label{asympt scalar}
\end{equation}%
\begin{equation}
\begin{array}{lll}
g_{rr}=\displaystyle\frac{l^{2}}{r^{2}}-\frac{4l^{2}\chi ^{2}}{r^{3}}%
+O(r^{-4}) &  & \displaystyle g_{tt}=-\frac{r^{2}}{l^{2}}+O(1) \\[2mm] 
g_{tr}=O(r^{-2}) &  & g_{\varphi \varphi }=r^{2}+O(1) \\[1mm] 
g_{\varphi r}=O(r^{-2}) &  & g_{t\varphi }=O(1)%
\end{array}
\label{asympt metric}
\end{equation}%
where $\chi =\chi (t,\varphi )$, and $\alpha $ is an arbitrary constant.
Remarkably, this set of asymptotic conditions is also left invariant under
the conformal group in two dimensions, spanned by two copies of the Virasoro
algebra. It was found that the effect of relaxing the asymptotic conditions
is such that the generators of the asymptotic symmetries acquire a
nontrivial contribution from the scalar field. Following the
Regge-Teitelboim approach \cite{Regge-Teitelboim}, the canonical generators
are given by%
\begin{equation}
Q(\xi )=\frac{1}{16\pi G}\!\int \!\! d\varphi \left\{ \frac{\xi ^{\bot }}{lr}\!\left(
(g_{\varphi \varphi }-r^{2})\!-\!2r^{2}(lg^{-1/2}-1)\right)\!+\!2\xi ^{\varphi }\pi
_{\ \varphi }^{r}\!+\!\xi ^{\bot }\frac{2r}{l}\!\left[ \phi ^{2}-2l\frac{\phi
\partial _{r}\phi }{\sqrt{g_{rr}}}\right] \right\},  \label{Q}
\end{equation}%
where the massless BTZ black hole has been chosen as reference background.
The Poisson brackets of these generators were shown to span two copies of
the Virasoro algebra with the standard central charges:%
\begin{equation}
c^{+}=c^{-}=c=\frac{3l}{2G}\ .
\end{equation}

\section{Microscopic entropy of the hairy black hole: Holographic puzzle and
its resolution}

\label{puzzle}

The existence of hairy black holes, for $\nu \geq -1$, that fit within a set
of asymptotic conditions being invariant under the conformal group at the
boundary, whose corresponding charge algebra acquires a nontrivial central
extension, makes compulsory wondering whether their entropy could be obtained
from the asymptotic growth of the number of states by means of Cardy formula 
\cite{Cardy}, as it is the case for the BTZ black hole \cite{Strominger}.
Once the hairy black hole entropy (\ref{Entropy}) is expressed in terms of
its mass,%
\begin{equation}
S=\pi l\Theta _{\nu }\sqrt{\frac{2M}{3G(1+\nu )}}\ ,  \label{S(M) hairy}
\end{equation}%
at a first glance, for a reader that is slightly familiarized with the
standard form of the Cardy formula, 
\begin{equation}
S=2\pi \sqrt{\frac{c^{+}}{6}\tilde{\Delta}^{+}}+2\pi \sqrt{\frac{c^{-}}{6}%
\tilde{\Delta}^{-}}\ ,  \label{Cardy simple}
\end{equation}%
where $\tilde{\Delta}^{\pm }=\frac{1}{2}(Ml\pm J)$ are the eigenvalues of
the left and right Virasoro operators, it doesn't seem so obvious how this
task could be successfully performed. In order to clarify the picture, some
remarks are worth to be pointed out:

{\Large $\cdot$} For a fixed value of the mass $M$, there are (at least) two
different static and circularly symmetric black holes; namely, the BTZ black
hole (in vacuum), and the hairy black hole (when the nontrivial scalar field
is switched on).

{\Large $\cdot$} Note that, since the hairy black hole depends on a single integration
constant, the scalar field cannot be switched off keeping the mass fixed.
This means that the hairy and BTZ black holes cannot be smoothly deformed
into each other, which suggests that they belong to different disconnected
sectors.

Thus, for a fixed value of the mass, despite the fact that there are (at
least) two different black hole configurations, Eq. (\ref{Cardy simple})
only reproduces the entropy of the BTZ black hole, which corresponds to the
vacuum sector. This raises a puzzle: How can the entropy of the hairy black
hole be obtained by means of Cardy formula?

As it is explained in section \ref{microscopic}, formula (\ref{Cardy simple}%
) implicitly assumes that the ground state is AdS spacetime. Thus, since the
entropy obtained from (\ref{Cardy simple}) does not reproduces (at least)
the sum of the entropy of the hairy and vacuum black holes, AdS spacetime
should be regarded as a suitable ground state \emph{only} for the vacuum sector.
Hence, if holography works, one should expect that the hairy sector
possesses a different ground state, disconnected from the vacuum one, so
that Cardy formula successfully reproduces the entropy of the hairy black
hole when this is taken into account.

The resolution of the puzzle comes from the fact that the suitable ground
state for the hairy sector indeed exists and it is described by a soliton.
The full picture can then be summarized as follows:

{\Large $\cdot$} The BTZ and the hairy black hole belong to disconnected sectors,
each with its own ground state. For the vacuum sector, the ground state
corresponds to AdS spacetime, while for the hairy sector, the ground state
corresponds to the scalar soliton.

{\Large $\cdot$} The soliton fulfills what is expected for a ground state: It is
smooth and regular everywhere, as well as devoid of integration constants.

{\Large $\cdot$} In analogy with what occurs for the vacuum, the energy spectrum of
the hairy sector consists of a continuous part bounded from below by zero
(hairy black holes), a gap (naked singularities), and a ground state with
negative mass fixed by the fundamental constants of the theory (soliton).

It is then reassuring to verify that Cardy formula reproduces the entropy of
the hairy black hole in exact agreement with the semiclassical result. This
is performed in section \ref{microscopic}. In what follows, the precise
analytic form of the scalar soliton is discussed.

\section{Scalar soliton}

\label{scalarSOL}

The field equations that correspond to the action (\ref{Action}) with the
self-interaction potential (\ref{Potential}), in the case of $\nu =0$ admit
the following exact solution:%
\begin{equation}
ds^{2}=l^{2}\left( -\frac{4(1+\rho ^{2})^{4}}{(3+2\rho ^{2})^{2}}d\tau ^{2}+%
\frac{64(1+\rho ^{2})^{3}}{(3+2\rho ^{2})^{4}}d\rho ^{2}+\frac{64}{81}\rho
^{2}(1+\rho ^{2})d\varphi ^{2}\right) \ ,  \label{soliton metric nu=0}
\end{equation}%
with%
\begin{equation}
\phi (\rho )=\mathrm{arctanh}\sqrt{\frac{1}{3+2\rho ^{2}}}\ ,
\label{soliton scalar field nu=0}
\end{equation}%
where the coordinates range according to $-\infty <\tau <\infty $, $0\leq
\rho <\infty $, and $0\leq \varphi <2\pi $. Note that the solution is devoid
of integration constants. It is also simple to verify that the soliton is
smooth everywhere and it fits within the relaxed set of asymptotically AdS
conditions given by Eq. (\ref{asympt metric}). This latter property can be
explicitly checked by performing a coordinate transformation defined through%
\begin{equation}
\tau =\frac{8t}{9l},\quad \rho =\sqrt{\frac{9}{8}\frac{r}{l}-\frac{1}{2%
}},  \label{coord nu 0}
\end{equation}%
with $r \geq \frac{4 l}{9}$, so that the metric and the scalar field read%
\begin{equation}
ds^{2}=-\frac{(4l+9r)^{4}}{81l^{2}(8l+9r)^{2}}dt^{2}+\frac{81l^{2}(4l+9r)^{3}%
}{(9r-4l)(8l+9r)^{4}}dr^{2}+\left( r^{2}-\frac{16}{81}l^{2}\right) d\varphi
^{2}\ ,  \label{soliton asympt}
\end{equation}%
\begin{equation}
\phi (r)=\mathrm{arctanh}\sqrt{\frac{4l}{8l+9r}}\ ,
\label{scalar field asymtp}
\end{equation}%
respectively.

In the case of $\nu >-1$, the solution generalizes as%
\begin{align}
\lefteqn{\!\!\!\!\!\!\!\!ds^{2}=l^{2}\left( 1+\frac{1}{\alpha _{\nu }(1+\rho
^{2})}\right) ^{-2}\times}  \notag \\
& \!\!\!\!\left( \!\!-(1+\rho ^{2})^{2}d\tau ^{2}+\frac{4d\rho ^{2}}{2+\rho
^{2}+\displaystyle\frac{c_{\nu }}{1+\rho ^{2}}}+\left( \frac{2}{2+c_{\nu }}%
\right) ^{2}\rho ^{2}\left( 2+\rho ^{2}+\frac{c_{\nu }}{1+\rho ^{2}}\right)
d\varphi ^{2}\!\!\right) \!,  \label{soliton with nu}
\end{align}%
with%
\begin{equation}
\phi (\rho )=\mathrm{arctanh}\,\sqrt{\frac{1}{1+\alpha _{\nu }(1+\rho ^{2})}}%
\ ,  \label{scalar field with nu}
\end{equation}%
and where\footnote{%
The previous case $\nu =0$ is recovered since $\Theta _{0}=\frac{4}{3}$%
, and thus\ $\alpha _{0}=2$ and $c_{0}=\frac{1}{4}$. For $-1<\nu <\infty $,
these constants range as $0<\alpha _{v}<\infty $, and $1>c_{\nu }>0$ (see
appendix \ref{a2}).}%
\begin{equation}
\alpha _{\nu }:=\frac{1}{2}\left( \Theta _{\nu }+\sqrt{\Theta _{\nu
}^{2}+4\Theta _{\nu }}\right) \ \mathrm{and\ }c_{\nu }:=\frac{2(1+\nu)}{\alpha
_\nu^3} \ ,  \label{alpha  and c}
\end{equation}%
just depend on the self-interaction parameter $\nu $, and $\Theta _{\nu }$
is defined in Eq. (\ref{Schuster}), so that there are no integration
constants. The soliton is regular and its causal structure coincides with
the one of AdS spacetime. Applying the following change of coordinates:%
\begin{equation}
\tau =\frac{\Theta _{\nu }\alpha _{\nu }}{3(1+\nu )}\frac{t}{l},\quad
\rho =\sqrt{\frac{3(1+\nu )}{\alpha _{\nu }\Theta _{\nu }}\frac{r}{l}+\frac{1%
}{\alpha _{\nu }}-1}\ ,  \label{coord nu}
\end{equation}%
one verifies that the soliton with a nonvanishing value of $\nu $ also
fulfills the set of relaxed asymptotically AdS conditions defined by (\ref%
{asympt metric}). These coordinates are useful in order to compute the mass
of the soliton, which can be readily obtained from the canonical generators
defined by the surface integrals in Eq. (\ref{Q}). The mass is found to be
given by%
\begin{equation}
M_{\mathrm{sol}}=Q(\partial _{t})=-\frac{\Theta _{\nu }^{2}}{24G(1+\nu )}\ ,
\label{Msol}
\end{equation}%
and as expected, depends only on the Newton constant and the self-interaction parameter. Note that the soliton mass is manifestly negative,
and it turns out to be bounded in between zero and the mass of AdS
spacetime, i.e. for the allowed values of the self-interaction parameter, $-1<\nu<\infty$, the soliton mass
ranges according to (see appendix \ref{a2})%
\begin{equation}
-\frac{1}{8G}<M_{\mathrm{sol}}<0\ .  \label{M sol range}
\end{equation}

\section{Microscopic entropy of the hairy black hole}

\label{microscopic}

As mentioned in section \ref{puzzle}, the entropy of the hairy black hole
can be microscopically computed provided the soliton is regarded as the
ground state of the hairy sector. In order to achieve this task, the
asymptotic growth of the number of states, given by Cardy formula, has to be
recalled from scratch.

If the spectrum of the Virasoro operators $L_{0}^{\pm }$, whose eigenvalues
are denoted by $\Delta ^{\pm }$, is such that the lowest eigenvalues $\Delta
_{0}^{\pm }$ are nonvanishing (i.e., for $\Delta _{0}^{\pm }\neq 0$), Cardy
formula reads \cite{Cardy, Carlip, MuInPark}%
\begin{equation}
S=2\pi \sqrt{\frac{\left( c^{+}-24\Delta _{0}^{+}\right) }{6}\left( \Delta
^{+}-\frac{c^{+}}{24}\right) }+2\pi \sqrt{\frac{\left( c^{-}-24\Delta
_{0}^{-}\right) }{6}\left( \Delta ^{-}-\frac{c^{-}}{24}\right) }\ ,
\label{Carlip reloaded}
\end{equation}%
where it is assumed that the ground state is non degenerate.

We would like pointing out here that, in terms of the shifted Virasoro
operators%
\begin{equation}
\tilde{L}_{0}^{\pm }:=L_{0}^{\pm }-\frac{c^{\pm }}{24}\ ,
\label{Shifted Virasoro}
\end{equation}%
this formula can be rewritten as follows:%
\begin{equation}
S=4\pi \sqrt{-\tilde{\Delta}_{0}^{+}\tilde{\Delta}^{+}}+4\pi \sqrt{-\tilde{%
\Delta}_{0}^{-}\tilde{\Delta}^{-}}\ ,  \label{Cardy super reloaded}
\end{equation}%
where ($\tilde{\Delta}_{0}^{\pm }$) $\tilde{\Delta}^{\pm }$ correspond to
the (lowest) eigenvalues of $\tilde{L}_{0}^{\pm }$. Thus, remarkably, the
asymptotic growth of the number of states can also obtained if one only
knows the spectrum of $\tilde{L}_{0}^{\pm }$ without making any explicit
reference to the central charges.

Note that unitarity, and the fact that expression (\ref{Cardy super reloaded}%
) makes sense only for negative lowest eigenvalues $\tilde{\Delta}_{0}^{\pm
} $, impose the following bounds\footnote{%
The lowest bounds hold provided $c^{\pm }>1$ (see e.g. \cite{Libros}).}%
\begin{equation}
-\frac{c^{\pm }}{24}\leq \tilde{\Delta}_{0}^{\pm }<0\text{\ .}
\label{Bound Delta zero}
\end{equation}

\begin{figure}[h!]
\centering
\includegraphics[scale=1.12]{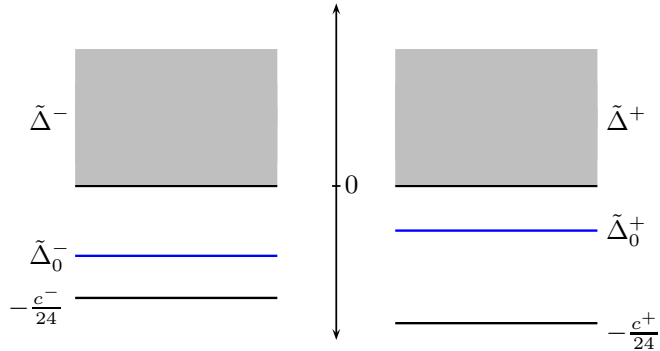}\caption{{\footnotesize The asymptotic growth of the number of states can be written exclusively in
terms of the spectrum of the shifted Virasoro operators $\tilde{L}_{0}^{\pm
} $. Then, it is given by $\rho (\tilde{\Delta}^{+},\tilde{\Delta}^{-})=\rho
(\tilde{\Delta}^{+})\rho (\tilde{\Delta}^{-})$, with $\rho (\tilde{\Delta}%
^{\pm })=\rho (\tilde{\Delta}_{0}^{\pm })\exp \left( 4\pi \sqrt{-\tilde{%
\Delta}_{0}^{\pm }\tilde{\Delta}^{\pm }}\right) $, where $\tilde{\Delta}%
_{0}^{\pm }$ and $\rho (\tilde{\Delta}_{0}^{\pm })$ correspond to the lowest
eigenvalues of $\tilde{L}_{0}^{\pm }$ and their degeneracies, respectively.}}%
\end{figure}

Formula (\ref{Cardy super reloaded}), or equivalently (\ref{Carlip reloaded}%
), provide the suitable ground in order to obtain the black hole entropy for
both sectors, namely vacuum and hairy, from the asymptotic number of states
in the microcanonical ensemble. This is performed assuming that the
eigenvalues of $\tilde{L}_{0}^{\pm}$ are given by their corresponding
canonical generators expressed by the surface integrals in Eq. (\ref{Q}),
which are related with the mass and the angular momentum according to%
\begin{equation}
\tilde{\Delta}^{\pm}=\frac{1}{2}(Ml\pm J)\ .  \label{Virasoros}
\end{equation}
Explicitly, this can be seen as follows:

\bigskip

\emph{Vacuum sector: }As it was explained in Sec. \ref{puzzle}, for the
vacuum sector, the ground state corresponds to AdS spacetime, whose lowest
eigenvalue is given by 
\begin{equation*}
\tilde{\Delta}_{0}^{\pm }=\frac{l}{2}M_{\mathrm{AdS}}=-\frac{l}{16G}\ .
\end{equation*}

Hence, since in the vacuum sector $\tilde{\Delta}_{0}^{\pm}$ can be
expressed in terms of the central charges, i.e., $\tilde{\Delta}_{0}^{\pm}=-%
\frac {c^{\pm}}{24}$, one verifies that formula (\ref{Cardy super reloaded})
reduces to (\ref{Cardy simple}), which by virtue of (\ref{Virasoros})
exactly reproduces the entropy of the BTZ black hole \cite{Strominger}.

\bigskip

\emph{Hairy sector:} In this case, the ground state is described by the
scalar soliton, for which

\begin{equation}
\tilde{\Delta}_{0}^{\pm }=\frac{l}{2}M_{\mathrm{sol}}=-\frac{l\Theta _{\nu
}^{2}}{48G(1+\nu )}\ .  \label{Delta0Sol}
\end{equation}%
Note that from Eq. (\ref{M sol range}), the lowest eigenvalues of the
Virasoro operators in (\ref{Delta0Sol}) fulfill the bounds given by Eq. (\ref%
{Bound Delta zero})\footnote{%
Indeed, in the semiclassical regime $c^{\pm }>>1$, since $l>>G$.}.
Therefore, since for the hairy black hole $\tilde{\Delta}^{\pm }=\frac{l}{2}%
M $, formula (\ref{Cardy super reloaded}), with (\ref{Delta0Sol}) gives 
\begin{equation*}
S=4\pi l\sqrt{-M_{\mathrm{sol}}M}=\pi l\Theta _{\nu }\sqrt{\frac{2M}{%
3G(1+\nu )}}\ ,
\end{equation*}%
in exact agreement with its semiclassical entropy (\ref{S(M) hairy}).

Thus, for fixed values of the global charges, it is confirmed that the total
entropy comes form the contribution of each sector.

\section{Discussion and comments}

\label{discussion}

Scalar solitons have been shown to exist in General Relativity with
minimally and conformally coupled self-interacting scalar field in three
dimensions. The soliton is regular everywhere, has a finite negative mass
and fulfills the same set of asymptotically AdS conditions as the hairy
black hole which are relaxed as compared with the ones of Brown-Henneaux. It
is worth pointing out that the soliton and the hairy black hole also obey
the same boundary conditions, since the constant $\alpha$ that appears in
Eq. (\ref{asympt scalar}) takes the same value, $\alpha=-\frac{2}{3}$, in
both cases.

The very existence of the scalar soliton appears to be essential in order to
reproduce the hairy black hole entropy from a microscopic counting. Indeed,
the soliton turns out to be the suitable nondegenerate ground state that is
required to obtain the entropy from the asymptotic growth of the number
states given by Cardy formula.

These results can also be extended for the (grand) canonical ensemble (see
appendix \ref{a3}), as well as for the rotating case by applying a boost in
the \textquotedblleft $t-\varphi $\textquotedblright\ cylinder.

The lack of integration constants for the soliton can also be understood
from the fact that the solution (\ref{soliton with nu}), (\ref{scalar field
with nu}) can be recovered from the hairy black hole (\ref%
{scalar bh}), (\ref{metric bh}) performing a \textquotedblleft double Wick
rotation\textquotedblright\ of the form $\varphi \rightarrow i\tau $, $%
t\rightarrow i\varphi $, with suitable rescalings and redefining the radial
coordinate\footnote{%
The following identity, $\Theta _{\nu }\alpha _{\nu }\left( 1+(1+\nu )\alpha
_{\nu }^{-3}\right) =3(1+\nu )$ also turns out to be useful.}. In fact, the
Euclidean black hole solution is regular provided the Euclidean time period
is fixed as the inverse of the Hawking temperature in Eq. (\ref{Temperature
and Mass}). Hence, in order to obtain a different Lorentzian solution
without closed timelike curves, the (former) angle has to be unwrapped, and
as a consequence the integration constant disappears by a simple rescaling.
Analogously, in the vacuum sector, one recovers AdS spacetime from the BTZ
black hole.

Note that since this construction is purely geometrical, it should apply
quite generically to obtain solitons from black holes in three dimensions%
\footnote{%
Indeed, as shown in \cite{OTT}, this mechanism successfully generates a
soliton from a black hole with ``gravitational hair" in vacuum within the context of the BHT
massive gravity theory \cite{BHT}, regardless the value of the cosmological
constant.}. This could also be interpreted as a manifestation of modular
invariance of the dual theory at the boundary.

As a final remark, one can note that, as it occurs for the BTZ black hole,
in the hairy sector there is also a gap between the ground state and the
hairy black holes, and thus it is natural to wonder whether the hairy black
hole could be obtained from some sort of identifications of the soliton.
Their possible local equivalence is suggested by the fact their scalar
invariants constructed out from the curvature and its derivatives coincide.
This can be easily seen as follows: Since the hairy black hole is static and
spherically symmetric, the invariants can only depend on $l$, $\nu$, and the
rescaled radial coordinate $\rho=r/B$; therefore, since the soliton is
obtained from this geometry by a double Wick rotation which only
interchanges the role of time and the angle, their scalar invariants are
necessarily the same. Nonetheless, it is simple to verify that the soliton
admits no additional Killing vectors apart from the manifest ones, generated
by $\partial_{t}$ and $\partial_{\phi}$, and hence the possible
identifications cannot be along an isometry.

\acknowledgments The authors thank Abhay Ashtekar, Fabrizio Canfora,
Nathalie Deruelle, Gast\'{o}n Giribet, Shinji Mukohyama, Alfredo P\'{e}rez, David Tempo and Jorge Zanelli for useful discussions. This work has been partially
funded by the following Fondecyt grants: 1085322, 1095098, 1100755, 3100123, and by the Conicyt grant \textquotedblleft Southern Theoretical Physics
Laboratory\textquotedblright\ ACT-91.
The Centro de Estudios Cient\'{\i}ficos (CECS) is funded by the Chilean
Government through the Centers of Excellence Base Financing Program of
Conicyt. CIN is funded by Conicyt and the
Gobierno Regional de Los R\'{\i}os.

\appendix

\section{Scalar soliton in the conformal frame}

\label{a1}

The self-interaction potential in Eq. (\ref{Potential}) has a natural
interpretation through the relation between the conformal and Einstein
frames. This can be seen performing a precise conformal transformation,
followed by a scalar field redefinition of the form 
\begin{equation}
\hat{g}_{\mu\nu}=\left( 1-\hat{\phi}^{2}\right) ^{-2}g_{\mu\nu }\quad %
\mbox{and} \quad \hat{\phi}=\tanh\left(\phi\right) \ ,  \label{map}
\end{equation}
so that the action given by (\ref{Action}) with (\ref{Potential}) reduces to
the one for General Relativity with cosmological constant and a conformally
coupled scalar field, given by 
\begin{equation}  \label{accionconformal}
I[\hat{g},\hat{\phi}]=\frac{1}{\pi G}\int d^{3}x\sqrt{-\hat{g}}\left( \frac{%
\hat{R}+2l^{-2}}{16}-\frac{1}{2}(\nabla\hat{\phi})^{2}-\frac {1}{16}\hat{R}%
\hat{\phi}^{2}+\frac{\nu}{8l^{2}}\hat{\phi}^{6}\right) \;.
\end{equation}
Note that in this frame, the self-interaction potential turns out to be
singled out requiring the matter piece of the action to be conformally
invariant; i.e., unchanged under local rescalings of the form $\hat{g}_{\mu
\nu}\rightarrow\lambda^{2}(x)\hat{g}_{\mu\nu}$, and $\hat{\phi}\rightarrow
\lambda^{-1/2}(x)\hat{\phi}$.

In the conformal frame the soliton acquires a simple form. In the case of $%
\nu =0$ the metric and the scalar field are given by 
\begin{eqnarray}
l^{-2}d\hat{s}^{2} &=&-(1+\rho ^{2})^{2}d\tau^{2}+\frac{16(1+\rho ^{2})}{%
(3+2\rho ^{2})^{2}}d\rho ^{2}+\left( \frac{4}{9}\right) ^{2}\rho ^{2}\frac{%
(3+2\rho ^{2})^{2}}{(1+\rho ^{2})}d\varphi ^{2},  \label{solitonconformenu0}
\\
\hat{\phi} &=&\sqrt{\frac{1}{3+2\rho ^{2}}},
\end{eqnarray}%
respectively; and when the self-interaction coupling is switched on, for $%
\nu >-1$, the solution generalizes according to 
\begin{eqnarray}
l^{-2}d\hat{s}^{2} &=&-(1+\rho ^{2})^{2}d\tau^{2}+\frac{4d\rho ^{2}}{2+\rho ^{2}+%
\frac{c_{\nu }}{1+\rho ^{2}}}+\left( \frac{2}{2+c_{\nu }}\right) ^{2}\rho
^{2}\left( 2+\rho ^{2}+\frac{c_{\nu }}{1+\rho ^{2}}\right) d\varphi ^{2},
\label{solitonconformenud0} \\
\hat{\phi} &=&\sqrt{\frac{1}{1+\alpha _{\nu }(1+\rho ^{2})}},
\end{eqnarray}%
where the constants $\alpha _{\nu }$ and $c_{\nu }$ are defined in Eq. (\ref%
{alpha and c}).

Note that the map between both frames (\ref{map}) is invertible along
the whole spacetime, since the conformal factor $(1-\hat{\phi}^{2})^{-2}$ is
positive. Thus, in the conformal frame the soliton is also smooth and
regular everywhere.

In the case of $\nu =0$, the corresponding black hole solution was found in 
\cite{Martinez:1996gn}, which extends for $\nu >-1$ as in Ref. \cite%
{HMTZ-2+1}.

\section{Ground state mass bounds}

\label{a2}

Here it is shown that a ground state of mass $M_{0}$ is bounded according to%
\begin{equation}
-\frac{1}{8G}\leq M_{0}<0\ ,  \label{Bound ground state}
\end{equation}%
which by virtue of (\ref{Delta0Sol}), agrees with the bound in Eq. (\ref%
{Bound Delta zero}) for the lowest eigenvalues of the shifted Virasoro
operators. The lowest bound is saturated in vacuum, since in this case $%
M_{0}=M_{\mathrm{AdS}}=-\frac{1}{8G}$. Besides, for the range of the self-interaction parameter $-1<\nu <\infty $, for which the soliton and the hairy
black hole exist, the mass of the soliton%
\begin{equation*}
M_{\mathrm{sol}}=-\frac{\Theta _{\nu }^{2}}{24G(1+\nu )}\ ,
\end{equation*}%
fulfills%
\begin{equation}
-\frac{1}{8G}<M_{\mathrm{sol}}<0\ .  \label{Bound2}
\end{equation}%
This can be seen as follows. The function $\Theta _{\nu }$ defined as 
\begin{equation}
\Theta _{\nu }=2(z\bar{z})^{\frac{2}{3}}\frac{z^{\frac{2}{3}}-\bar{z}^{\frac{%
2}{3}}}{z-\bar{z}}\ ,\quad \mbox{with}\quad z=1+i\sqrt{\nu }\ ,
\end{equation}%
is monotonically increasing. It vanishes for $\nu =-1$ according to 
\begin{equation}
\Theta _{\nu }\xrightarrow[\nu\rightarrow{}-1]\,{2^{2/3}(1+\nu )^{2/3}-\frac{%
(1+\nu )^{4/3}}{2^{2/3}}+\mathcal{O}\left( (1+\nu )^{5/3}\right) \ },
\end{equation}%
and then grows so that $\Theta _{0}=\frac{4}{3}$, while for large $\nu $,
asymptotically behaves as 
\begin{equation}
\Theta _{\nu }\xrightarrow[\nu\rightarrow{}\infty]\,{\sqrt{3\nu }-\frac{2}{3}%
+\mathcal{O}\left( \nu ^{-1/2}\right) \ }.
\end{equation}%
Therefore, the behavior of the soliton mass reads%
\begin{equation}
GM_{\mathrm{sol}}=\left\{ 
\begin{array}{ccc}
\displaystyle-2^{-2/3}\frac{(1+\nu )^{1/3}}{6}+\frac{1+\nu }{12}+\mathcal{O}%
\left( (1+\nu )^{4/3}\right) & : & \nu \rightarrow -1\ , \\ 
\displaystyle-\frac{1}{8}+\frac{\nu ^{-1/2}}{6\sqrt{3}}+\mathcal{O}\left(
\nu ^{-1}\right) & : & \nu \rightarrow \infty \ ,%
\end{array}%
\right.
\end{equation}%
so that the bound (\ref{Bound2}) is fulfilled.

\section{Cardy formula and black hole entropy in the (grand) canonical
ensemble}

\label{a3}

Conformal symmetry at the boundary is described by two commuting copies of
the Virasoro algebra, so that left and right movers are decoupled, and hence
they can be at equilibrium at different temperatures $T_{\pm}=\beta_{%
\pm}^{-1}$. Therefore, the total free energy is given by the sum of each of
their free energies in the canonical ensemble, i.e.,%
\begin{equation}
F=\left( \beta_{+}\tilde{\Delta}^{+}+\beta_{-}\tilde{\Delta}^{-}\right)
l^{-1}-S\ ,  \label{F+-}
\end{equation}
where $S$ is given by (\ref{Cardy super reloaded}). Therefore, at the
equilibrium the entropy can be written in terms of the lowest eigenvalues of
the shifted Virasoro operators and the temperatures of left and right movers
according to%
\begin{equation}
S=-8\pi^{2}l\left( \tilde{\Delta}_{0}^{+}T_{+}+\tilde{\Delta}%
_{0}^{-}T_{-}\right) \ .  \label{Cardy Canonical super reloaded}
\end{equation}
In turn, from the knowledge of the entropy and left and right temperatures,
what one extracts from (\ref{Cardy Canonical super reloaded}) are the lowest
eigenvalues of the shifted Virasoro operators instead of the central
charges. Indeed, by virtue of Eq. (\ref{Shifted Virasoro}), this formula is
equivalently expressed in terms of the \textquotedblleft effective central
charges" $c_{eff}^{\pm}:=c^{\pm}-24\Delta_{0}^{\pm}$ as follows\footnote{%
This might be related to the discrepancy in the central charges found in the
context of warped AdS black holes \cite{Warped-AdS, Compere-Detournay}.}%
\begin{equation}
S=\frac{\pi^{2}}{3}\left( c_{eff}^{+}T_{+}+c_{eff}^{-}T_{-}\right) \ .
\label{Cardy canonical reloaded}
\end{equation}
Formula (\ref{Cardy Canonical super reloaded}), or equivalently (\ref{Cardy
canonical reloaded}), also allows to obtain the black hole entropy for the vacuum
and hairy sectors in the canonical ensemble. Indeed, since for the black
holes the free energy is given by%
\begin{equation*}
F=\beta M+\beta \Omega _{+}J-S\ ,
\end{equation*}%
where $\beta =T^{-1}$, and $\Omega _{+}$ stands for the angular velocity of
the horizon, then from Eqs. (\ref{F+-}) and (\ref{Virasoros}), the
corresponding left and right temperatures are found to be 
\begin{equation}
T_{\pm }=\frac{T}{1\pm l\Omega _{+}}\ .  \label{T+-}
\end{equation}%
Hence, the black hole entropy for the vacuum and hairy sectors can be
recovered from (\ref{Cardy Canonical super reloaded}) with (\ref{T+-}) as
follows:

\emph{Vacuum sector:} In this case the ground state corresponds to AdS
spacetime, whose lowest eigenvalue is given by 
\begin{equation*}
\tilde{\Delta}_{0}^{\pm }=\frac{l}{2}M_{\mathrm{AdS}}=-\frac{l}{16G}=-\frac{%
c^{\pm }}{24}\ .
\end{equation*}%
Thus, (\ref{Cardy Canonical super reloaded}) reduces to 
\begin{equation*}
S=\frac{\pi ^{2}}{3}\left( c^{+}T_{+}+c^{-}T_{-}\right) \ ,
\end{equation*}%
which exactly reproduces the entropy of the BTZ black hole \cite{Strominger}.

\bigskip

\emph{Hairy sector:} The ground state is described by the scalar soliton,
for which

\begin{equation}
\tilde{\Delta}_{0}^{\pm }=\frac{l}{2}M_{\mathrm{sol}}=-\frac{l\Theta _{\nu
}^{2}}{48G(1+\nu )}\ .  \label{Delta0Sol2}
\end{equation}%
In the absence of angular momentum, by virtue of (\ref{Temperature and Mass}%
), left and right temperatures coincide, i.e., $T_{\pm }=T$. Therefore, Eq. (%
\ref{Cardy Canonical super reloaded}) reduces to%
\begin{equation}
S=-8\pi ^{2}l^{2}\ M_{\mathrm{sol}}T=\pi l\Theta _{\nu }\sqrt{\frac{2M}{%
3G(1+\nu )}}\ ,
\end{equation}%
in exact agreement with the semiclassical entropy given by (\ref{S(M) hairy}).


\begin{thebibliography}{99}
\bibitem{Breitenlohner-Freedmann} P.~Breitenlohner and D.~Z.~Freedman,
Phys.\ Lett.\ B \textbf{115}, 197
(1982); 
Annals Phys.\ \textbf{144}, 249 (1982). 

\bibitem{Mezin-Townsend} L.~Mezincescu and P.~K.~Townsend, 
Annals Phys.\ \textbf{160}, 406 (1985). 

\bibitem{HMTZ-2+1} M.~Henneaux, C.~Mart\'{\i}nez, R.~Troncoso and J.~Zanelli, 
Phys.\ Rev.\ D \textbf{65}, 104007 (2002) [arXiv:hep-th/0201170]. 

\bibitem{HMTZ-Log} M.~Henneaux, C.~Mart\'{\i}nez, R.~Troncoso and J.~Zanelli, 
Phys.\ Rev.\ D \textbf{70}, 044034 (2004) [arXiv:hep-th/0404236]. 

\bibitem{HMTZ-D} M.~Henneaux, C.~Mart\'{\i}nez, R.~Troncoso and J.~Zanelli, 
Annals Phys.\ \textbf{322}, 824 (2007) [arXiv:hep-th/0603185]. 

\bibitem{Hertog-Maeda} T.~Hertog and K.~Maeda, 
JHEP \textbf{0407}, 051 (2004)
[arXiv:hep-th/0404261]. 

\bibitem{Amsel-Marolf} A.~J.~Amsel and D.~Marolf, 
Phys.\ Rev.\ D \textbf{74},
064006 (2006) [Erratum-ibid.\ D \textbf{75}, 029901 (2007)]
[arXiv:hep-th/0605101]. 

\bibitem{Brown-Henneaux} J.~D.~Brown and M.~Henneaux, 
Commun.\ Math.\
Phys.\ \textbf{104}, 207 (1986). 

\bibitem{Henneaux-Teitelboim} M.~Henneaux and C.~Teitelboim,
Commun.\ Math.\ Phys.\ \textbf{98}, 391 (1985). 

\bibitem{Henneaux-D} M.~Henneaux, 
\emph{\textquotedblleft Asymptotically
Anti-De Sitter Universes In D = 3, 4 And Higher Dimensions\textquotedblright }, 
Proceedings of the Fourth Marcel Grossmann Meeting on General Relativity,
Rome 1985. R. Ruffini (Ed.), Elsevier Science Publishers B.V., pp. 959-966. 

\bibitem{MTZ1} C.~Mart\'{\i}nez, R.~Troncoso and J.~Zanelli, 
Phys.\ Rev.\ D \textbf{70}, 084035 (2004) [arXiv:hep-th/0406111]. 

\bibitem{MTZ2} C.~Mart\'{\i}nez, J.~P.~Staforelli and R.~Troncoso, 
Phys.\ Rev.\ D \textbf{74}, 044028 (2006) [arXiv:hep-th/0512022]. 

\bibitem{MTZ3} C.~Mart\'{\i}nez and R.~Troncoso, 
Phys.\ Rev.\ D \textbf{74}, 064007 (2006) [arXiv:hep-th/0606130]. 

\bibitem{MTZ4} C.~Charmousis, T.~Kolyvaris and E.~Papantonopoulos, 
Class.\ Quant.\ Grav.\ \textbf{26}, 175012 (2009) [arXiv:0906.5568 [gr-qc]]. 

\bibitem{MTZ5} A.~Anabal\'on and H.~Maeda, 
Phys.\ Rev.\ D \textbf{81}, 041501 (2010) [arXiv:0907.0219 [hep-th]]. 

\bibitem{MTZ6} T.~Kolyvaris, G.~Koutsoumbas, E.~Papantonopoulos and
G.~Siopsis, 
``A New Class of Exact Hairy Black Hole Solutions,''
arXiv:0911.1711 [hep-th]. 

\bibitem{N1} T.~Torii, K.~Maeda and M.~Narita, 
Phys.\ Rev.\ D \textbf{64}, 044007 (2001). 

\bibitem{N2} E.~Winstanley, 
Found.\ Phys.\ \textbf{33}, 111 (2003) [arXiv:gr-qc/0205092]. 

\bibitem{N3} T.~Hertog and G.~T.~Horowitz, 
Phys.\ Rev.\ Lett.\ \textbf{94}, 221301 (2005) [arXiv:hep-th/0412169]. 

\bibitem{N4} E.~Radu and E.~Winstanley, 
Phys.\ Rev.\ D \textbf{72}, 024017 (2005) [arXiv:gr-qc/0503095]. 

\bibitem{N5} D.~Hosler and E.~Winstanley,
  Phys.\ Rev.\  D {\bf 80}, 104010 (2009)
  [arXiv:0907.1487 [gr-qc]].
  
\bibitem{N6}  O.~J.~C.~Dias, R.~Monteiro, H.~S.~Reall and J.~E.~Santos,
 ``A scalar field condensation instability of rotating anti-de Sitter black holes,''
  arXiv:1007.3745 [hep-th].

\bibitem{HC}
  S.~S.~Gubser,
  Phys.\ Rev.\  D {\bf 78}, 065034 (2008)
 [arXiv:0801.2977 [hep-th]];
  S.~A.~Hartnoll, C.~P.~Herzog and G.~T.~Horowitz,
  Phys.\ Rev.\ Lett.\  {\bf 101}, 031601 (2008)
 [arXiv:0803.3295 [hep-th]].

\bibitem{reviews} S.~A.~Hartnoll, 
Class.\ Quant.\ Grav.\ \textbf{26}, 224002 (2009) [arXiv:0903.3246 [hep-th]];
C.~P.~Herzog,
  J.\ Phys.\ A  {\bf 42}, 343001 (2009)
[arXiv:0904.1975 [hep-th]];
G.~T.~Horowitz, ``Introduction to Holographic Superconductors,''
arXiv:1002.1722 [hep-th]. 


\bibitem{Manga} S.~S.~Gubser and A.~Nellore, 
JHEP \textbf{0904}, 008 (2009) [arXiv:0810.4554 [hep-th]]; 
P.~Basu, J.~Bhattacharya, S.~Bhattacharyya, R.~Loganayagam, S.~Minwalla
and V.~Umesh, 
``Small Hairy Black Holes in Global AdS Spacetime,''
arXiv:1003.3232 [hep-th]. 

\bibitem{BTZ} M.~Ba\~{n}ados, C.~Teitelboim and J.~Zanelli, 
Phys.\ Rev.\ Lett.\ \textbf{69}, 1849
(1992) [arXiv:hep-th/9204099]. 

\bibitem{BHTZ} M.~Ba\~{n}ados, M.~Henneaux, C.~Teitelboim and J.~Zanelli,
Phys.\ Rev.\ D \textbf{48}, 1506 (1993) [arXiv:gr-qc/9302012]. 

\bibitem{glenn} G.~Barnich, 
``Conserved charges in gravitational theories: Contribution from scalar fields,''
[arXiv:gr-qc/0211031]. 

\bibitem{glenn2} G.~Barnich, 
Class.\ Quant.\ Grav.\ \textbf{20}, 3685 (2003) [arXiv:hep-th/0301039]. 

\bibitem{GMT} J.~Gegenberg, C.~Mart\'{\i}nez and R.~Troncoso, 
Phys.\ Rev.\ D \textbf{67}, 084007 (2003) [arXiv:hep-th/0301190].

\bibitem{Clement} G.~Cl\'ement, 
Phys.\ Rev.\ D \textbf{68}, 024032 (2003) [arXiv:gr-qc/0301129]. 

\bibitem{BanadosThiesen} M.~Ba\~{n}ados and S.~Theisen, 
Phys.\ Rev.\ D \textbf{72}, 064019 (2005) [arXiv:hep-th/0506025]. 

\bibitem{Regge-Teitelboim} T.~Regge and C.~Teitelboim, 
Annals Phys.\ \textbf{88}, 286 (1974). 

\bibitem{Cardy} J.~L.~Cardy, 
Nucl.\ Phys.\ B \textbf{270}, 186 (1986). 

\bibitem{Strominger} A.~Strominger,  
JHEP \textbf{9802}, 009 (1998)  [arXiv:hep-th/9712251].  

\bibitem{Carlip} S.~Carlip, 
Class.\ Quant.\ Grav.\ \textbf{16}, 3327 (1999) [arXiv:gr-qc/9906126]. 

\bibitem{MuInPark} M.~I.~Park, 
Phys.\ Lett.\ B \textbf{597}, 237 (2004) [arXiv:hep-th/0403089]. 

\bibitem{Libros} P.~Di Francesco, P.~Mathieu and D.~Senechal, 
\textit{Conformal Field Theory},
New York, USA: Springer (1997).

\bibitem{OTT} J.~Oliva, D.~Tempo and R.~Troncoso, 
JHEP \textbf{0907}, 011 (2009) [arXiv:0905.1545 [hep-th]]. 

\bibitem{BHT} E.~A.~Bergshoeff, O.~Hohm and P.~K.~Townsend, 
Phys.\ Rev.\ Lett.\ \textbf{102}, 201301 (2009) [arXiv:0901.1766 [hep-th]]. 

\bibitem{Martinez:1996gn} C.~Mart\'{\i}nez and J.~Zanelli, 
Phys.\ Rev.\ D \textbf{54} 3830 (1996) [arXiv:gr-qc/9604021].


\bibitem{Warped-AdS} D.~Anninos, W.~Li, M.~Padi, W.~Song and A.~Strominger, 
JHEP \textbf{0903}, 130 (2009) [arXiv:0807.3040 [hep-th]]. 

\bibitem{Compere-Detournay} G.~Comp\`{e}re and S.~Detournay, 
Class.\ Quant.\ Grav.\ \textbf{26}, 012001 (2009) [Erratum-ibid.\ \textbf{26}%
, 139801 (2009)] [arXiv:0808.1911 [hep-th]]. 
\end{thebibliography}
\end{document}